\documentclass[journal]{IEEEtran}
\usepackage{amsmath,amsfonts}
\usepackage{algorithmic}
\usepackage{algorithm}
\usepackage{array}
\usepackage[caption=false,font=normalsize,labelfont=sf,textfont=sf]{subfig}
\usepackage{textcomp}
\usepackage{stfloats}
\usepackage{url}
\usepackage{verbatim}
\usepackage{graphicx}
\usepackage{cite}
\usepackage{amsmath}
\usepackage{xcolor}
\usepackage{graphicx}
\usepackage[utf8]{inputenc} 
\usepackage{etex}
\usepackage[utf8]{inputenc}
\usepackage{amsfonts}
\usepackage{amsmath}
\usepackage{amssymb}
\usepackage{booktabs}
\usepackage{caption}
\usepackage{ctable}
\usepackage{enumitem}
\usepackage{mathptmx}
\usepackage{mathtools}
\usepackage{multicol}
\usepackage[normalem]{ulem}
\usepackage{setspace}
\usepackage{tabularx}
\usepackage{textcomp}
\usepackage[english]{babel}
\usepackage{geometry}
\usepackage[normalem]{ulem}

\newcommand{\expr}[1]{\text{exp}\left( #1 \right)}
\newcommand{\exps}[1]{\text{exp}\left[#1 \right]}

\newcommand{\qv}{\ensuremath{{\bf q}}}

\newcommand{\QMF}{{\ensuremath{Q_{MF}}}}
\newcommand{\QMD}{{\ensuremath{Q_{MD}}}}

\newcommand{\rv}{\ensuremath{\bf{ r}}}

\newcommand{\rvz}{\ensuremath{{\bf r}_0}}

\title{Limitations to Electrical Probing of Spontaneous Polarization in Ferroelectric-Dielectric Heterostructures}

\author{Mattia Segatto ~\IEEEmembership{Graduate Student Member,~IEEE,} Riccardo Fontanini ~\IEEEmembership{Graduate Student Member,~IEEE,} Francesco Driussi ~\IEEEmembership{Member,~IEEE,} Daniel Lizzit ~\IEEEmembership{Member,~IEEE,} David Esseni ~\IEEEmembership{Fellow,~IEEE}}

\begin{document}

\markboth{Journal of Electron Devices Society, ~Vol., No., December~2021}%
{}


\onecolumn

\maketitle
\vspace{-0.7truecm}

© 2022 IEEE.  Personal use of this material is permitted.  Permission from IEEE must be obtained for all other uses, in any current or future media, including reprinting/republishing this material for advertising or promotional purposes, creating new collective works, for resale or redistribution to servers or lists, or reuse of any copyrighted component of this work in other works.

\begin{abstract}
An accurate estimate of the ferroelectric polarization in ferroelectric-dielectric stacks is important from a materials science perspective, and it is also crucial for the development of ferroelectric based electron devices. This paper revisits the theory and application of the PUND technique in Metal-Ferroelectric-Dielectric-Metal (MFDM) structures by using analytical derivations and numerical simulations. In an MFDM structure the results of the PUND technique may largely differ from the polarization actually switched in the stack, which in turn is different from the remnant polarization of the underlying ferroelectric. The main hindrances that prevent PUND measurements from providing a good estimate of the polarization switching in MFDM stacks are thus discussed. The inspection of the involved physical quantities, not always accessible in experiments, provides a useful insight about the main sources of the errors in the PUND technique, and clarifies the delicate interplay between the depolarization field and the charge injection and trapping in MFDM stacks with a thin dielectric layer.
\end{abstract}


\section{Introduction}
\label{Sec:Intro}

Thanks to the discovery of a robust ferroelectricity in hafnium oxide thin films,$^{\text{\cite{Boscke_APL2011}}}$ several intriguing applications of ferroelectricity in CMOS electron devices have been proposed and are being presently scrutinized. Device concepts include nanoscale CMOS FETs exploitating the effective negative capacitance to improve the subthreshold\IEEEpubidadjcol swing,$^{\text{\cite{Salahuddin_NL2008,Jain_TED2014, Khan_TED2016, Rollo_EDL2018, Rollo_EDL2018_trap, Pentapati_TED2020}}}$ as well as Ferroelectric Tunnelling Junctions (FTJs),$^{\text{\cite{Max_JEDS2019,Fontanini_ESSDERC2021}}}$ and ferroelectric FETs,$^{\text{\cite{Mulaosmanovic_EDL2020,Lizzit_ESSDERC2021}}}$ which may be used as non--volatile  memories or as memristors for neuromorphic computing applications.$^{\text{\cite{Slesazeck_Nanotechnology2019}}}$

\begin{figure}[t]
\centering
\includegraphics[width=0.5\textwidth]{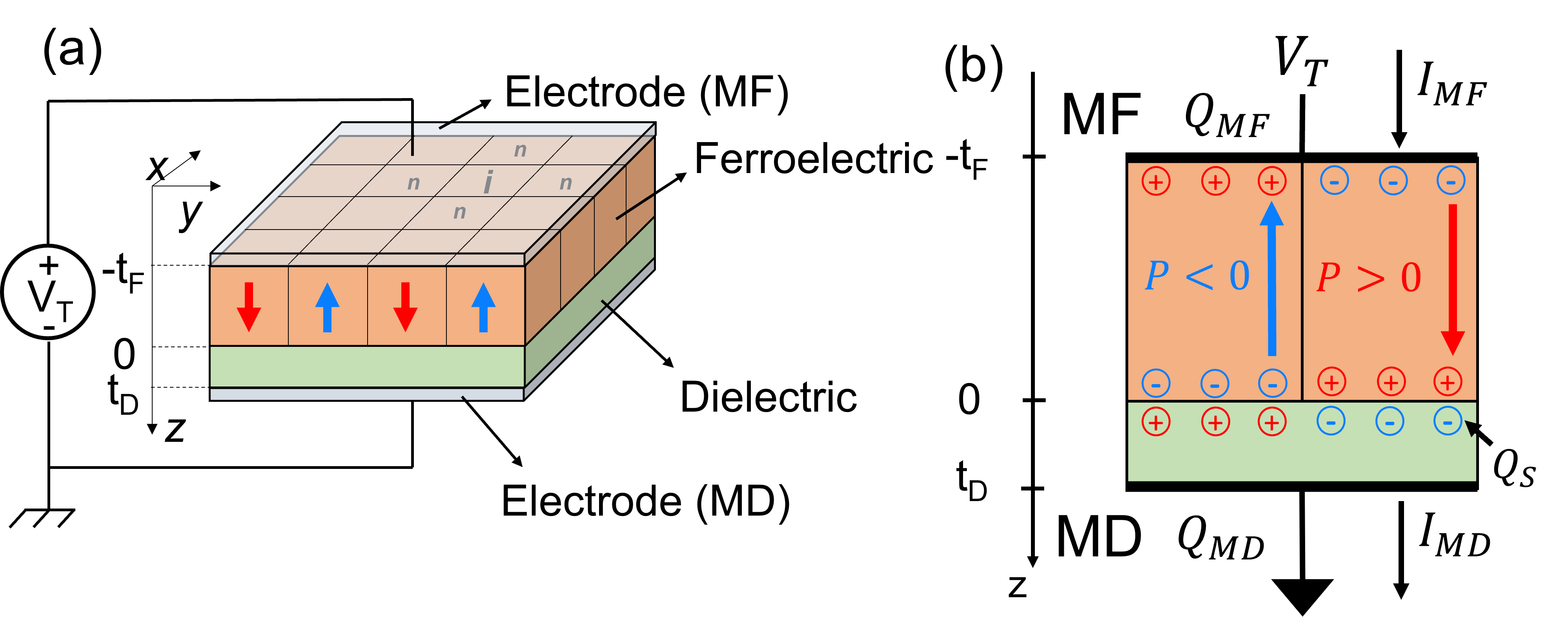}
\caption{\footnotesize Ferroelectric capacitors and related symbols. a) Three dimensional sketch of an MFDM capacitor, where $t_F$ and $t_D$ are the ferroelectric and dielectric thicknesses, and $V_T$ is the external bias. b) Cross section of an MFDM structure. The polarization, $P$, is taken positive when it points to the dielectric. $Q_{MF}$, $I_{MF}$, $Q_{MD}$, $I_{MD}$ denote  the charges and currents at the MF and MD electrodes, respectively, whereas $Q_{S}$ is the charge trapped at the ferroelectric-dielectric interface.
}
\vspace{-0.4truecm}
\label{Fig:MFIM_Sketches}
\end{figure}

\begin{figure}[htbp]
\centering
\includegraphics[width=0.8\linewidth]{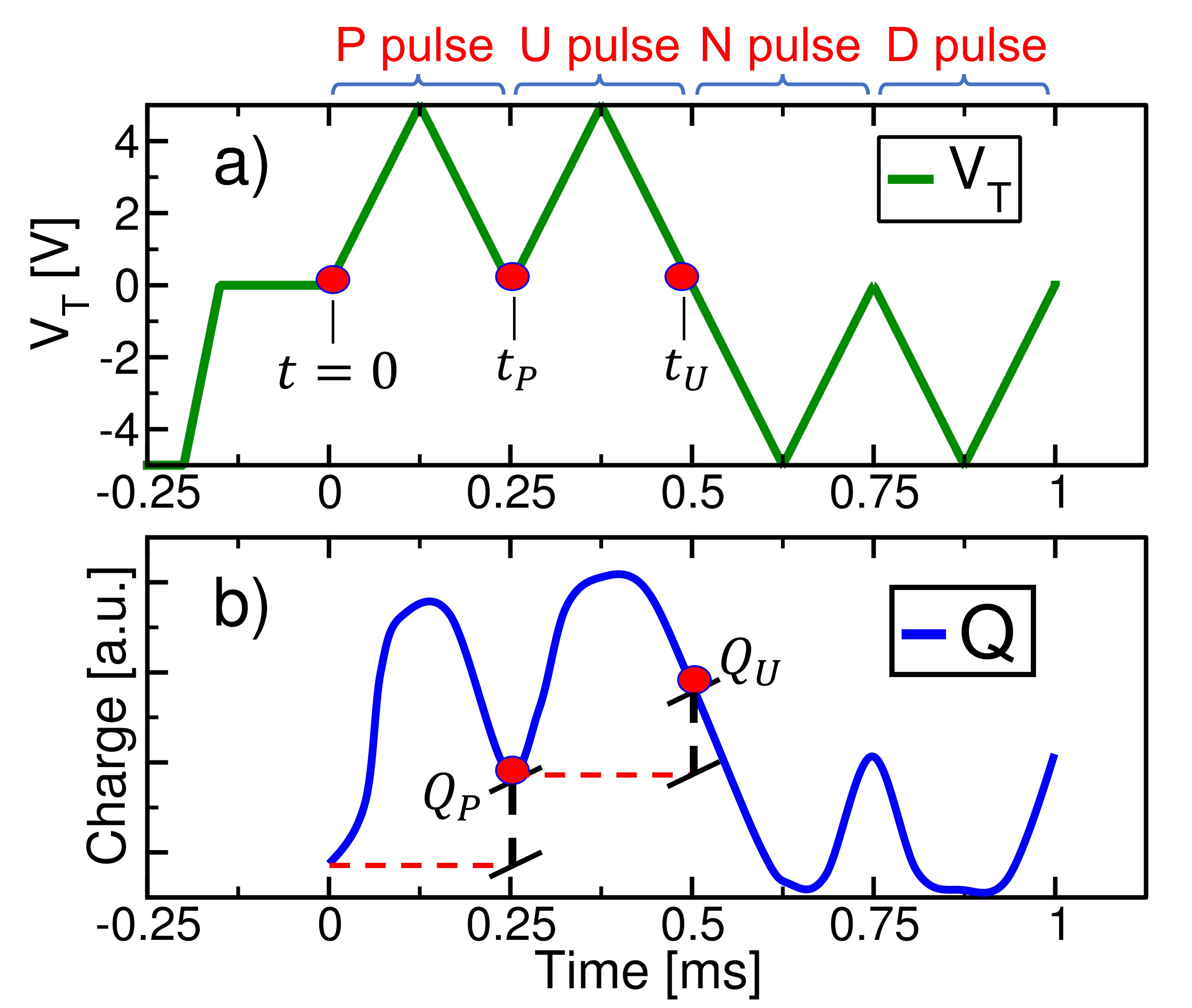}
\caption{\footnotesize a) Examples of the $V_T$ waveform used in PUND measurements. The 250 $\mu$s pulse width has been used in all the simulation results reported in Section \ref{Sec:Numerical_Sims} (so $t_U$ = $2t_P$), if not otherwise stated. A preset pulse at $V_T$ = $-5$ V and for 125 $\mu$s is used to set an initial negative polarization state; b) Charge waveform sketch during the PUND simulation. Some key points are defined in order to simplify the notation of the paper.}
\vspace{-2mm}
\label{Fig:PUND_sketch}
\end{figure}

In most of the above material systems and electron devices, one of the materials adjacent to the ferroelectric is a dielectric (see also Figure \ref{Fig:MFIM_Sketches}), and the working principle of the devices relies on the influence that the ferroelectric polarization, $P$, exerts on the band bending inside the device.
Quite understandably, a dependable determination of $P$ is of primary importance in ferroelectric materials and ferroelectric based electron devices. To this purpose the Positive-Up-Negative-Down (PUND) measurement technique was originally conceived for Metal-Ferroelectric-Metal (MFM) structures$^{\text{\cite{Rabe_book, Fukunaga_JPSJ2008}}}$ (see also PUND waveforms in Figure \ref{Fig:PUND_sketch}), and it is still routinely used also in Metal-Ferroelectric-Dielectric-Metal (MFDM) device structures.$^{\text{\cite{Lomenzo_AVS2014, Mikheev_ACS2019, Ryu_Nature2019, Deshpande_SD2021}}}$
The main goal of the PUND technique for an MFM structure is an accurate determination of the difference, 2$P_r$, between the polarization at zero ferroelectric field for the positive and negative polarization state. In an MFM stack with ideal metal electrodes the field can be zeroed by using a zero external voltage, and the PUND measurements essentially intend to minimize the contributions to 2$P_r$ due to the background ferroelectric polarization and to possible leakage currents (see also the discussion in Section \ref{Sec:General_Model}). \IEEEpubidadjcol In an MFDM structure, however, a zero external voltage cannot force a zero ferroelectric field due to the depolarization field, and
the application of the PUND technique to MFDM devices is not straightforward in several respects. In fact the interpretation of PUND results in an MFDM stack can lead to artifacts and to a misleading information  about the polarization of the underlying ferroelectric layer.
We here revisit the theory and application of PUND measurements in MFDM structures, and to this purpose we use both analytical derivations and a comprehensive modelling framework,
that has been previously validated and calibrated against experiments. Our results show that: a) the determination of the spontaneous ferroelectric polarization is challenging in MFDM structures even if the charge injection and trapping in the dielectric stack is negligible;
b) in the presence of a non negligible charge injection and trapping, the variations of spontaneous polarization and trapped charge are inextricably entangled, which further complicates the extraction of the switched polarization;
c) the $t_D$ dependence of the extracted polarization can be an artifact due to the ($t_D$ dependent) interplay between the depolarization field and charge trapping.
The paper is organized as follows. In Section \ref{Sec:General_Model} we present the theoretical background behind the extraction of the spontaneous polarization from the terminal currents measured by the PUND technique. In Section \ref{Sec:Numerical_Sims} we provide an overview of our {\it in house} modelling framework and then in Secs. \ref{Sec:Results} and \ref{Sec:Errors} we report simulation results about the ability of the PUND technique to determine the switched polarization in MFDM structures, and offer an insight about the main sources of errors. In Section \ref{Sec:Conclusions} we propose a few concluding remarks and in the Appendix provide additional details about some aspects of the modelling framework.

\section{Charge and current at the electrodes in the MFDM structure}
\label{Sec:General_Model}

Let us consider the MFDM structure sketched in Figure \ref{Fig:MFIM_Sketches}(a), where $\rv$$=$$(x,y)$ and $z$ are the coordinate in the plane of the ferroelectric-dielectric interface and in the direction normal to the interface, while
\QMF, \QMD\ are the charges per unit area respectively at the MF and MD electrodes, respectively (see Figure \ref{Fig:MFIM_Sketches}).
Assuming a perfect screening in the metal electrodes, the electrostatic problem is linear and \QMF, \QMD\ can be written by using appropriate Green's functions for the charges in the structure.
More specifically, for \QMF\ we have
\begin{equation}
Q_{MF}(t) = \dfrac{1}{A}\int\displaylimits_A P(\textbf{r},t)\, d\textbf{r}  + \dfrac{1}{A}\int\displaylimits_A \varepsilon_0\varepsilon_{F} \, E_{FT}(\textbf{r},t)  d\textbf{r}
\label{Eq:QMF}
\end{equation}
%
where $A$ is the device area, $P$ is the ferroelectric spontaneous polarization,
$E_{FT}(\textbf{r}, t)$ denotes the $z$ component of the electric field at the position $\textbf{r}$ of the MF-FE interface (namely at $z$$=$$-t_F$).

\begin{figure*}[ptb]
	\begin{minipage}[t]{0.5\textwidth}
		\centering
		\includegraphics[width=0.85\textwidth]{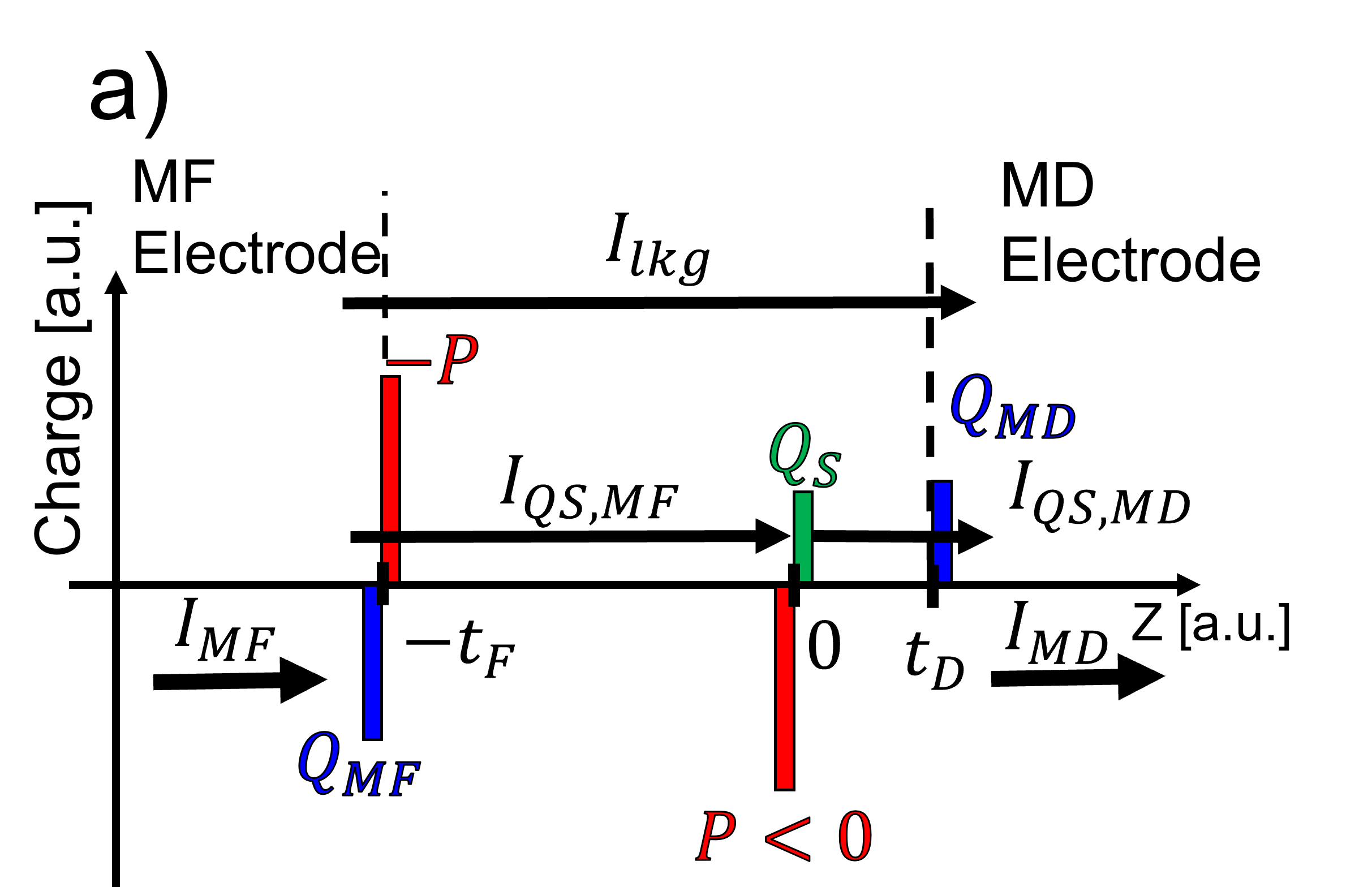}
	\end{minipage}
	\begin{minipage}[t]{0.4\textwidth}
		\centering
		\includegraphics[width=0.9\textwidth]{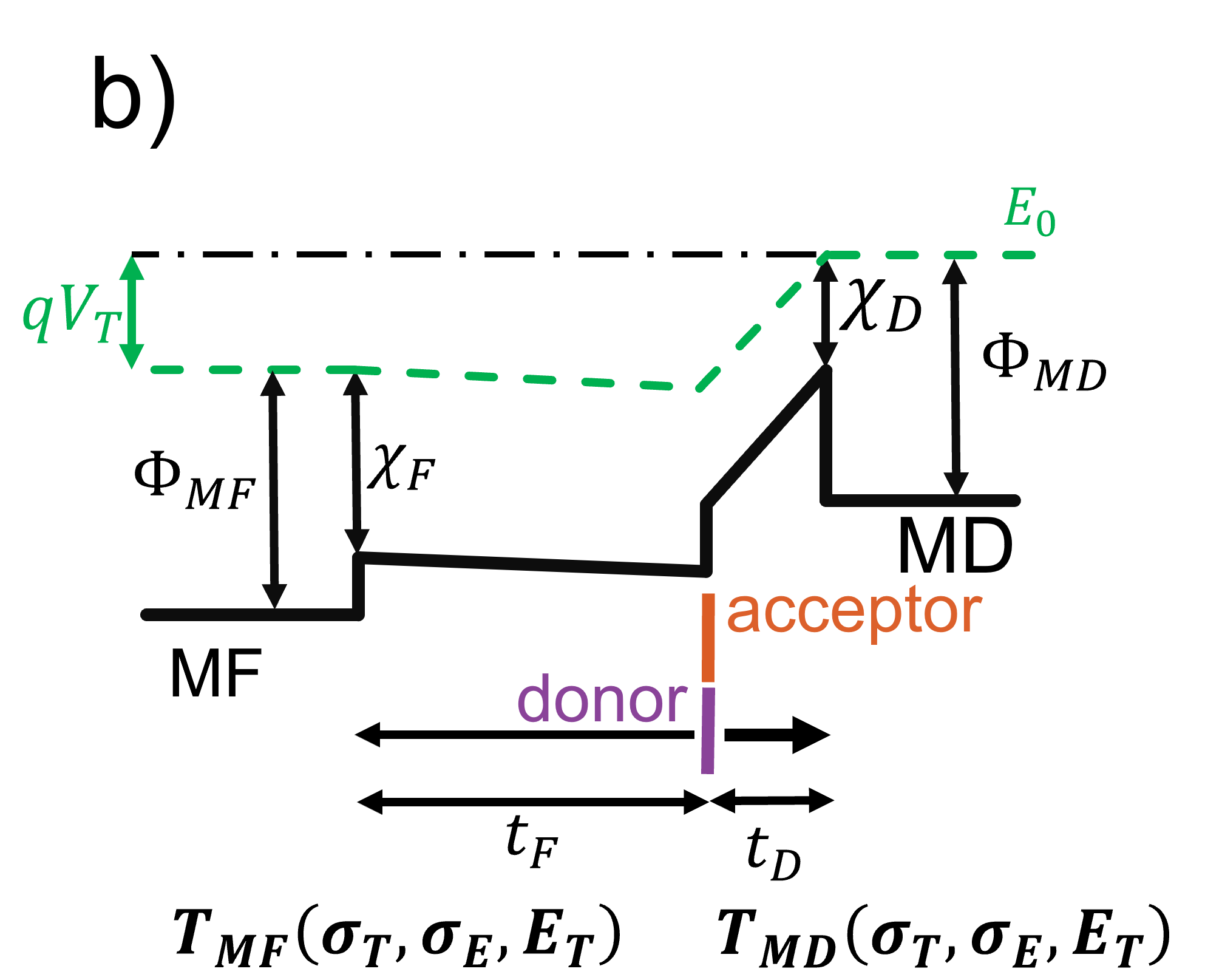}
	\end{minipage}
	\caption{\footnotesize a) Sketch of the sheet charges in the MFDM structure of Figure \ref{Fig:MFIM_Sketches}(b), where $Q_{MF}$, $Q_{MD}$ denote the overall charges at the electrodes. The picture illustrates an example corresponding to a negative polarization $P$ and a positive interface charge $Q_S$. $I_{QS,MF}$ and $I_{QS,MD}$ denote the currents due to trapping and detrapping at the FE-DE interface, and $I_{lkg}$ is a possible leakage current through the whole structure. $I_{MF}$, $I_{MD}$ denote the overall currents at the electrodes (see Equation \ref{Eq:IMF_DEF}), that are used in the PUND characterization technique. b) Sketch of the band diagram in an MFDM stack, where $\chi_D$, $\chi_F$ are the electron affinity of the dielectric and ferroelectric material, while $\Phi_{MD}$, $\Phi_{MF}$ are the workfunctions of the MD and MF electrodes (that are equal in the simulations of this work, Table \ref{Tab:Sims_Param}). The energy position of acceptor and donor type traps is also depicted. The tunnelling coefficients $T_{MD}$ and $T_{MF}$ depend on the energy, $\sigma_\text{E}$, and geometric, $\sigma_\text{T}$, cross sections of the traps, as well as on the traps energy $E_T$.
	}
	\label{Fig:Charges_Currents}
	\vspace{-5mm}
\end{figure*}

At any time $t$, the $E_{FT}(\textbf{r},t)$ is determined by the external bias $V_T$ and by the charges in the dielectric stack, that are charges located at coordinate $\textbf{r}_0$ defined as $\textbf{r}$ at $z$ = 0. In this latter respect, we here define the depolarization field, $E_{DP}(\textbf{r},t)$, at the MF-FE interface as the field produced by the distribution of the total charge $[P(\textbf{r}_0,t)$ + $Q_S(\textbf{r}_0,t)]$ at the FE-DE interface, where $Q_S(\textbf{r}_0,t)$ is an interface charge due to fixed Coulomb centers or to interface traps. We also similarly introduce $E_{\rho}(\textbf{r},t)$ as the field produced by the remaining charge densities $\rho (\textbf{r}_0,z_0,t)$ in the dielectric stack at $z_0 \ne 0$. The $E_{FT}(\textbf{r},t)$ can thus be written as
\begin{equation}
\varepsilon_0\varepsilon_{F} \, E_{FT}(\textbf{r},t) = C_S \, V_T + \varepsilon_0\varepsilon_{F} [ \, E_{DP}(\textbf{r},t) + E_{\rho}(\textbf{r},t) \,]
\label{Eq:epsF_EFT}
\end{equation}
where $C_S$$=$$(1/C_D+1/C_F)^{-1}$, with $C_D$$=$$\varepsilon_0\varepsilon_{D}/t_D$, $C_F$$=$$\varepsilon_0\varepsilon_{F}/t_F$ and $t_D$, $\varepsilon_{D}$ being the thickness and relative permittivity of the dielectric, and $t_F$, $\varepsilon_{F}$ being the thickness and background permittivity of the ferroelectric (see Figure \ref{Fig:MFIM_Sketches}(b)).
By substituting Equation \ref{Eq:epsF_EFT} in Equation \ref{Eq:QMF} we obtain
\begin{equation}
	\begin{split}
		Q_{MF}(t) = &C_S \, V_T(t) + P_{AV}(t)+\\
		&+ \varepsilon_0 \varepsilon_{F} \, E_{DP,AV}(t) + \varepsilon_0\varepsilon_{F} \, E_{\rho,AV}(t)
	\end{split}
\label{Eq:QMF_Edep}
\end{equation}
where $P_{AV}$, $E_{DP,AV}$ and $E_{\rho,AV}$ denote respectively the average polarization and average fields at the MF-FE interface.
From \ref{Sec:SUPPL_Eq_Pav} of the Appendix it can be inferred that the average $E_{DP}$ and $E_{\rho}$ can be written as:
\begin{subequations}
\begin{equation}
\varepsilon_0\varepsilon_{F} \, E_{DP,AV} = \dfrac{1}{A}\int\displaylimits_A [P(\textbf{r}_0) + Q_S(\textbf{r}_0)] \, G_{MF}(\textbf{r}_0,z_0) \, d\textbf{r}_0
\vspace{-0.2truecm}
\label{Eq:Edep}
\end{equation}
\begin{equation}
\varepsilon_0\varepsilon_{F} \, E_{\rho,AV} = \dfrac{1}{A}\int\displaylimits_A\int\displaylimits_{-t_F}^{t_D} \rho (\textbf{r}_0,z_0) \, G_{MF}(\textbf{r}_0,z_0)\, dz_0\, d\textbf{r}_0
\label{Eq:Erho}
\end{equation}
\label{Eq:EdepErho}
\end{subequations}
where $G_{MF}(\textbf{r}_0,z_0)$ is the Green's function
defined as
\begin{equation}
G_{MF}(\textbf{r}_0,z_0) = \dfrac{\varepsilon_0 \varepsilon_F}{e} \int\displaylimits_A E_{FT}\left[\textbf{r}_0,z_0\right](\textbf{r})\,d \textbf{r}  \, \, .
\label{Eq:GMF_DEF}
\end{equation}
with $E_{FT}\left[\textbf{r}_0,z_0\right](\textbf{r})$ being the field $E_{FT}(\textbf{r})$ produced by a point charge $e$ located at $(\textbf{r}_0,z_0)$.
In \ref{Sec:Qmd_suppl} of Appendix we also demonstrate that, under realistic assumptions, the $G_{MF}(\textbf{r}_0,z_0)$ for the MFDM structure in Figure \ref{Fig:MFIM_Sketches}(b) is independent of $\textbf{r}_0$ and it can be evaluated analytically. For a charge located at the FE-DE interface, for example, we have $G_{MF}(\textbf{r}_0,0)$$\simeq$$-(C_F/C_0)$, which allows to rewrite Equation \ref{Eq:Edep} as
\begin{equation}
\varepsilon_0\varepsilon_{F} \, E_{DP,AV} \simeq - \dfrac{C_F}{C_0} ( P_{AV} +  Q_{S,AV} ) 
\label{Eq:Edep_Pav_QSav}
\end{equation}
At any $z_0 \ne 0$ the $G_{MF}(\textbf{r}_0,z_0)$ can be similarly expressed with a $z_0$ dependent capacitance ratio.
We now recall that PUND measurements are based on the integral of the transient current at the electrodes, hence by definition the experiments can probe only the variations of the polarization and charges in the device stack.
Here below the discussion is carried out in terms of the current $I_{MF}$ at the MF terminal; in \ref{Sec:Qmd_suppl} of Appendix we report the corresponding expression for $I_{MD}$.
In the presence of a trapping distributed throughout the device, it is thus difficult to express the influence of $\rho (\textbf{r}_0,z_0,t)$ on $I_{MF}$, because Equation \ref{Eq:Erho} shows that the information about the distribution along $z_0$ is required.
Consequently, hereafter we simplify the picture and assume that the time derivative of the charge trapped in the dielectric stack is dominated by the $(\partial Q_S$$/$$\partial t)$ term due to traps at the FE-DE interface, which implies\\ $\partial E_{\rho,AV}/\partial t$$\ll$$\partial E_{DP,AV}/\partial t$. We also assume that $Q_S(t)$ can change only through the terminal currents $I_{QS,MF}$, $I_{QS,MD}$ shown in Figure \ref{Fig:Charges_Currents}, and we let $I_{lkg}$ denote a possible leakage current, not contributing to trapping. The current at the MF electrode can thus be written as
\begin{equation}
	\begin{split}
		I_{MF} = &\dfrac{\partial Q_{MF}}{\partial t} + I_{QS,MF} + I_{lkg} = C_S \dfrac{\partial V_{T}}{\partial t} + 	\dfrac{\partial P_{AV}}{\partial t} \\
		&+ \varepsilon_0 \varepsilon_{F} \, \dfrac{\partial E_{DP,AV}}{\partial t} + I_{QS,MF} + I_{lkg} = \\
		= &C_S\dfrac{\partial V_T}{\partial t} + \dfrac{C_D}{C_0} \dfrac{\partial P_{AV}}{\partial t} - \dfrac{C_F}{C_0} \dfrac{\partial Q_{S,AV}}{\partial t} + I_{QS,MF} + I_{lkg}
		\label{Eq:IMF_DEF}
	\end{split}
\end{equation}
where $Q_{MF}$ has been expressed via Equation \ref{Eq:QMF_Edep} assuming $\partial E_{\rho,AV}/\partial t$$\ll$$\partial E_{DP,AV}/\partial t$, and in the last equality we have used Equation \ref{Eq:Edep_Pav_QSav}.\\
If we now consider the Positive (P) pulse of a PUND experiment starting at $t$ = 0 s with $V_T(0)$ =  0 V (see the waveform in Figure \ref{Fig:PUND_sketch}(a)), we can evaluate the charge $Q_P(t)$ (with $0 \leq t \leq t_P$) by integrating the expression for $I_{MF}$ in Equation \ref{Eq:IMF_DEF} and obtain
\begin{equation}
	\begin{split}
		Q_P(t) = &\int_0^t\, I_{MF}(t')\, dt' \approx \\
		 &\approx C_S V_T(t) + P_{AV}(t) + \varepsilon_0 \varepsilon_{F} E_{DP,AV}(t)+\\
		 &+ Q_{QS,MF}(t) + Q_{lkg}(t)
	\end{split}
\label{Eq:QP}
\end{equation}
where $Q_{QS,MF}$ and $Q_{lkg}$ are the integral of $I_{QS,MF}$ and $I_{lkg}$, respectively. Moreover it is understood that $P_{AV}$, $E_{DP,AV}$ and $Q_{QS,MF}$ in Equation \ref{Eq:QP} denote the variations from the corresponding values at $t$$=$0 or, equivalently, that Equation \ref{Eq:QP} conventionally assumes $P_{AV}$$=$$E_{DP,AV}$$=$$Q_{S,MF}$$=$$Q_{MF}$$=$0 at $t$$=$0. The charges $Q_U(t)$, $Q_N(t)$, $Q_D(t)$ during respectively the Up (U), Negative (N) and Down (D) pulses of the PUND technique have expressions equivalent to Equation \ref{Eq:QP}.\\
The first and last term at the right hand side of Equation \ref{Eq:QP} are the contributions due to, respectively, the linear polarization of the dielectrics and the leakage. Even in an MFM structure these contributions complicate the extraction of the remnant polarization $2P_{r}$$=$$P_{AV}(t=t_P)$, and PUND measurements address this issue by subtracting from $Q_P$ the charge $Q_U$ during the U pulse at the same external bias $V_T$.
In our theoretical framework such an approach results in the charge
$Q_{PU}$$=$$(Q_P$$-$$Q_U)$, that can be written as
\begin{equation}
	\begin{split}
	Q_{PU} \approx  &P_{AV}^{(P)} - P_{AV}^{(U)} +  \varepsilon_0\varepsilon_{F} \left ( E_{DP,AV}^{(P)} - E_{DP,AV}^{(U)} \right ) +\\
	&+ Q_{QS,MF}^{(P)}-Q_{QS,MF}^{(U)} +  Q_{lkg}^{(P)} - Q_{lkg}^{(U)}
	\end{split}
\label{Eq:QP_QU}
\end{equation}
where the apices $(P)$, $(U)$ identify the P and U pulse and all charges are evaluated at times corresponding to the same  $V_T$ value during either a rising or a falling $V_T$ ramp.
We recall that, as already mentioned about Equation\ref{Eq:QP},
the $P_{AV}$, $E_{DP,AV}$ and $Q_{QS,MF}$ in Equation \ref{Eq:QP_QU} denote the variations from the corresponding values at the beginning of either the P or the U pulse.

The second and third term in the right hand side of Equation \ref{Eq:QP_QU} are due respectively to the depolarization field and the current at the MF electrode contributing to trapping at the FE-DE interface.
In an MFM structure both these terms are negligible and, moreover, it is typically assumed that the polarization can be stabilized after the P pulse, so that $P_{AV}^{(U)}$ is much smaller than $P_{AV}^{(P)}$. Furthermore, it is also usually assumed that the leakage affects the measurements to a similar extent during the P and U pulse, leading to $Q_{lkg}^{(U)}$$\approx$$Q_{lkg}^{(P)}$.$^{\text{\cite{Schenk_TUFF2015}}}$ Under these circumstances Equation \ref{Eq:QP_QU} shows that the $Q_{PU}$ in an MFM stack can be interpreted as the $P_{AV}^{(P)}$ that we wish to determine.
In an MFDM structure, instead, the terms in Equation \ref{Eq:QP_QU} due to the depolarization field can be comparable to $P_{AV}^{(P)}$, and the term $(Q_{QS,MF}^{(P)}-Q_{QS,MF}^{(U)})$ may also give a sizeable contribution to $Q_{PU}$. Hence in an MFDM stack the interpretation of $Q_{PU}$ and the determination of $P_{AV}^{(P)}$ appear much more delicate that in the MFM counterpart. This is systematically investigated in Section\ref{Sec:Numerical_Sims} by using numerical simulations.

\section{Numerical simulations and discussion}
\label{Sec:Numerical_Sims}

The goal of the PUND measurements is to accurately determine the spontaneous polarization switched by the external bias, which is very challenging in an MFDM structure due to the depolarization field and the possible charge trapping. In this section, we use numerical simulations to investigate the possible errors and artifacts produced by PUND measurements in MFDM structures, and to provide some useful physical insights.

\subsection{Modelling framework and validation}
\label{Sec:Model}
Our in house developed simulation framework comprises models for the ferroelectric dynamics, a dynamic equation for the traps at the FE-DE interface and a description of a tunnelling injection from the MF and MD electrodes to the traps.
The dynamics of the ferroelectric domains is described by a formulation of the multi-domain Landau-Ginzburg-Devonshire (LGD) model more thoroughly discussed in $^{\text{\cite{Rollo_Nanoscale2020}}}$
\begin{equation}
\begin{split}
	t_F\,\rho\,\dfrac{\partial P_i}{\partial t} = &-\left(2\alpha_i\,P_i + 4\beta_i\,P_i^3 + 6\gamma_i\,P_i^5\right)t_F +\\
	&- \dfrac{t_F\, k}{d\, w}\sum_n\left(P_i-P_n\right)+\\
	&-\dfrac{1}{2}\sum_{j=1}^{n_D}\left( \dfrac{1}{C_{i,j}}+\dfrac{1}{C_{j,i}} \right)\, P_j +\dfrac{C_D}{C_0}V_T
\end{split}
\label{eq:MFIM_dPdt}
\end{equation}
where $\alpha$, $\beta$, $\gamma$ are the anisotropy constants, $\rho$ denotes the resistivity that sets a time scale $t_{\rho}$$=$$\rho/(2|\alpha |)$ for the ferroelectric switching. Moreover, the parameters $1/C_{i,j}$ describe the depolarization energy and depolarization field in the MFDM structure, while $k$ and $w$ are the coupling constant and the domain wall width involved in the formulation of the domain wall energy.
The mean values for $\alpha$, $\beta$, $\gamma$ used in simulations are reported in Table \ref{Tab:Sims_Param} and are referred to Hafnium-Zirconium-Oxide (HZO) ferroelectric  which is the current state-of-the art ferrolectric used in FTJs.
In all simulations the domain wall coupling $k$ was set to zero, by following recent first principles calculations for HfO$_2$.$^{\text{\cite{Lee_Science_2020}}}$ Moreover, we assumed a resistivity $\rho$$=$115 $\Omega$m, which is consistent with recently reported values for Hafnium-Zirconium (HZO) based capacitors,$^{\text{\cite{Kobayashi_IEDM2016, Kim_IEDM2020}}}$ and results in a time scale for the ferroelectric dynamics $t_{\rho}$$\approx$ $119.8$ ns. All simulations include a $n_D$$=$ 1024 and a domain size d=5nm; we verified that results are insensitive to a further increase of $n_D$.
The modelling for the ferroelectric dynamics has been extensively compared to transient negative capacitance measurements, demonstrating a good agreement with experiments in asymmetric MFDM dielectric stacks,$^{\text{\cite{Rollo_Nanoscale2020,Esseni_Nanoscale2021}}}$ and also in symmetric MFDFM structures.$^{\text{\cite{Hoffmann_AdvFunctionalMaterials_2021}}}$
The charge trapping model follows a first-order dynamic equation for the occupation $f_T$ of either acceptor or donor type traps at the FE-DE interface. By denoting with $c_{MD0}$, $c_{MF0}$ the capture rate from the metal MD and MF electrodes the equation governing $f_T$ can be written as
\begin{equation}
\dfrac{\partial f_T}{\partial t} = c_{MD0} \, [ \, f_{0,MD} - f_T \, ] + c_{MF0} \, [ \, f_{0,MF} - f_T \, ] \, \,
\vspace{-0.08truecm}
\label{eq:fT_Dyn}
\end{equation}
where $f_{0,M}(E_T)$$=$$1/[1+\text{exp}((E_T - E_{f,M})/(K_BT) )]$ is the Fermi occupation function in the metal electrodes, with $E_{f,MF}$=$E_{f,MD}-qV_B$. In the derivation of Equation \ref{eq:fT_Dyn} we used a detailed balance condition, ensuring that the steady state $f_T$ value at the equilibrium (i.e. for $V_T$=$0$ V) is given by the Fermi function. In this work, the capture rates were attributed to tunnelling from and to the electrodes, and the tunnelling transmission described according to a WKB approximation that involves the tunnelling effective mass in the two dielectrics $m_D$, $m_F$, and the area and energy cross sections $\sigma_T$ [m$^2$], $\sigma_E$ [eV] (see also Figure \ref{Fig:Charges_Currents}). More details about the trapping and tunnelling models may be found in$^{\text{\cite{Fontanini_ESSDERC2021, Fontanini_JEDS2021}}}$ reporting also a good agreement with the polarization versus voltage curves in FTJs with an MFDM structure and a thin dielectric layers. The values of $m_D$, $m_F$, $\sigma_T$, $\sigma_E$ used in this work are reported in Table \ref{Tab:Sims_Param}.\\
From the occupations $f_T$ one can readily calculate the charges $Q_{acc}$ and $Q_{don}$ respectively in acceptor and donor type traps, and finally the overall interface trapped charge $Q_S$$=$($Q_{acc}$$+$$Q_{don})$. The knowledge of the time dependent polarization and trapped charge, in turn, allows one to numerically calculate all the quantities discussed in Section \ref{Sec:General_Model}, such as $I_{MF}$, $P_{AV}$, $Q_{S,AV}$, and also $I_{MF,QS}$, $Q_{MF,QS}$.
\begin{table*}[!t]
\caption{\footnotesize \label{Tab:Sims_Param}Material parameters used in simulations for the	Hf$_{0.5}$Zr$_{0.5}$O$_2$$-$Al$_2$O$_3$ MFDM system. Here $\alpha$, $\beta$ and $\gamma$ are the mean values of the anisotropic constants, and calculations include domain to domain variations of the $\alpha_i$, $\beta_i$, $\gamma_i$ parameters (with $i$=$1,\, 2, \ldots, n_D$) corresponding to a ratio $\sigma_{EC}=10$\% between the standard deviation and the mean value of the coercive field $E_C$. m$_\text{D}$ and m$_\text{F}$ are the effective tunnelling masses for respectively Al$_2$O$_3$ and HZO, while $\sigma_\text{E}$, $\sigma_\text{T}$ denote respectively the energy and geometric cross section of the traps, that are used in the tunnelling model. The maximum energy values for the traps are $0.6$ and $1.3$ eV below the conduction band minimum at the FE-DE interface for acceptor and donor traps, respectively. Both traps type extend in energy for $2$ eV below their maximum.
	The electron affinity was set to $\chi_D$$=$ 1.4 eV for Al$_2$O$_3$$^{\text{\cite{Arya_JTAC2015}}}$ and to $\chi_F$$=$ 2.4 eV for HZO,$^{\text{\cite{Zheng_JPC2005}}}$ while the workfunction for both TiN metal electrodes was taken as $\Phi_M$$=$ 4.5 eV.$^{\text{\cite{Vitale_TED2010}}}$}
\centering
\begin{tabular}{|c| |c| |c| |c| |c| |c| |c| |c| |c|}
	\hline
	$\alpha$ [m/F]&$\beta$ [m$^{5}$/C$^{2}$/F] &$\gamma$ [m$^{9}$/C$^{4}$/F] & $\epsilon_F$ & $\epsilon_D$ & m$_\text{F}$ [m$_0$] &  m$_\text{D}$ [m$_0$] &$\sigma_\text{T}$ [m$^2$] & $\sigma_\text{E} $ [eV]\\
	\hline
	$-4.8\cdot10^8$
	& $\quad 1.46\cdot10^{9}$
	& $\quad 3.14\cdot 10^{10}$
	& $34$
	& $10$
	& $0.4$
	& $0.18$
	& $1\cdot 10^{-19}$
	& $7\cdot10^{-3}$ \\
	\hline
\end{tabular}
\end{table*}
\vspace{-0.2truecm}
\subsection{Simulations results}
\label{Sec:Results}
We used numerical simulations to emulate PUND measurements with a 1 kHz waveform and a 5 V peak voltage in an MFDM structure with different dielectric thicknesses $t_D$ and trap densities $N_{acc}$ and $N_{don}$. The ferroelectric HZO layer is 10 nm thick in all simulations.

We do not account for a possible leakage current flowing directly from MD to MF.
Here we did not attempt to model the leakage current because we think that leakage is strongly technology dependent, frequently governed by Poole-Frenkel and hopping mechanisms in the the HZO and, as such, difficult to describe in simulations.$^{\text{\cite{Schroeder_JAP2015}}}$ Moreover, while it is understood that a failure of the condition $Q_{lkg}^{(U)}$$\approx$$Q_{lkg}^{(P)}$ can induce artifacts in PUND experiments even in an MFM stack, this issue goes beyond the scope of the present paper, that is focused on the influence that the depolarization field and charge trapping have on the results of PUND measurements in MFDM structures.

\begin{figure}[tb]
\centering
\includegraphics[width=1\linewidth]{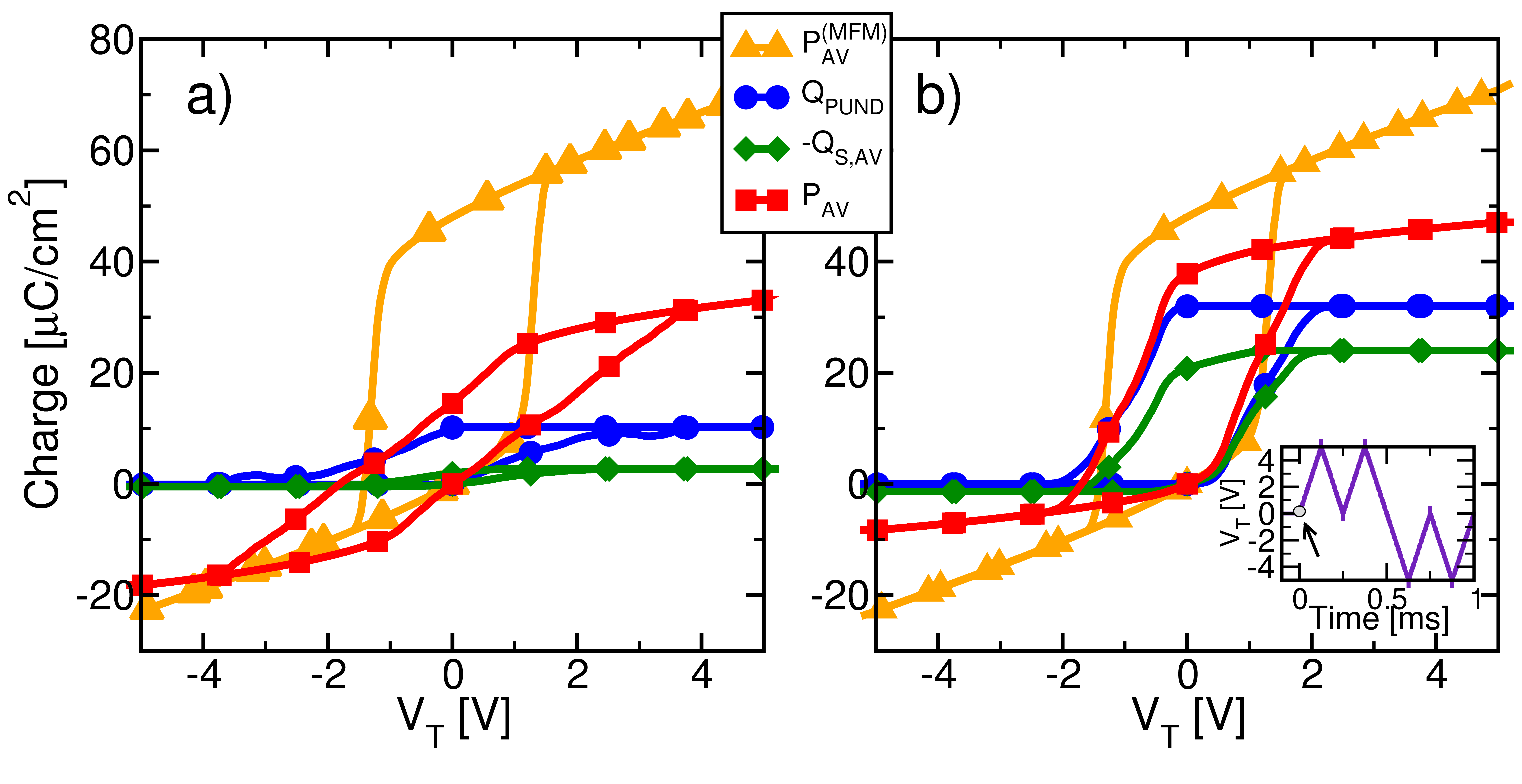}
\vspace{-0.3truecm}
\caption{\footnotesize Simulated charges corresponding to a 1 kHz PUND waveform applied to an MFDM structure. The $Q_{PUND}$ is either $Q_{PU}$ = $(Q_P$ - $Q_U)$ or $Q_{ND}$ = $(Q_N$ - $Q_D)$, respectively during the P pulse (i.e. for a positive $V_T$) or during the N pulse (i.e. for a negative $V_T$). The average polarization $P_{AV}$ and trapped charge $-Q_{S,AV}$ during the P and N pulses are also shown.	The Al$_2$O$_3$ layer thickness is $t_D$$=$1.5 nm. a) Results for acceptor and donor type trap densities $N_{acc}$ = $N_{don}$ = $0.5$ $\times$ $10^{13}$ [cm$^{-2}$eV$^{-1}$]; b) Results for $N_{acc}$$=$$N_{don}$$=$$4\times 10^{13}$ [cm$^{-2}$eV$^{-1}$].
}
\label{fig:pund_nacc}
\vspace{-0.6truecm}
\end{figure}

\begin{figure}[htb]
	\centering
	\includegraphics[width=1\linewidth]{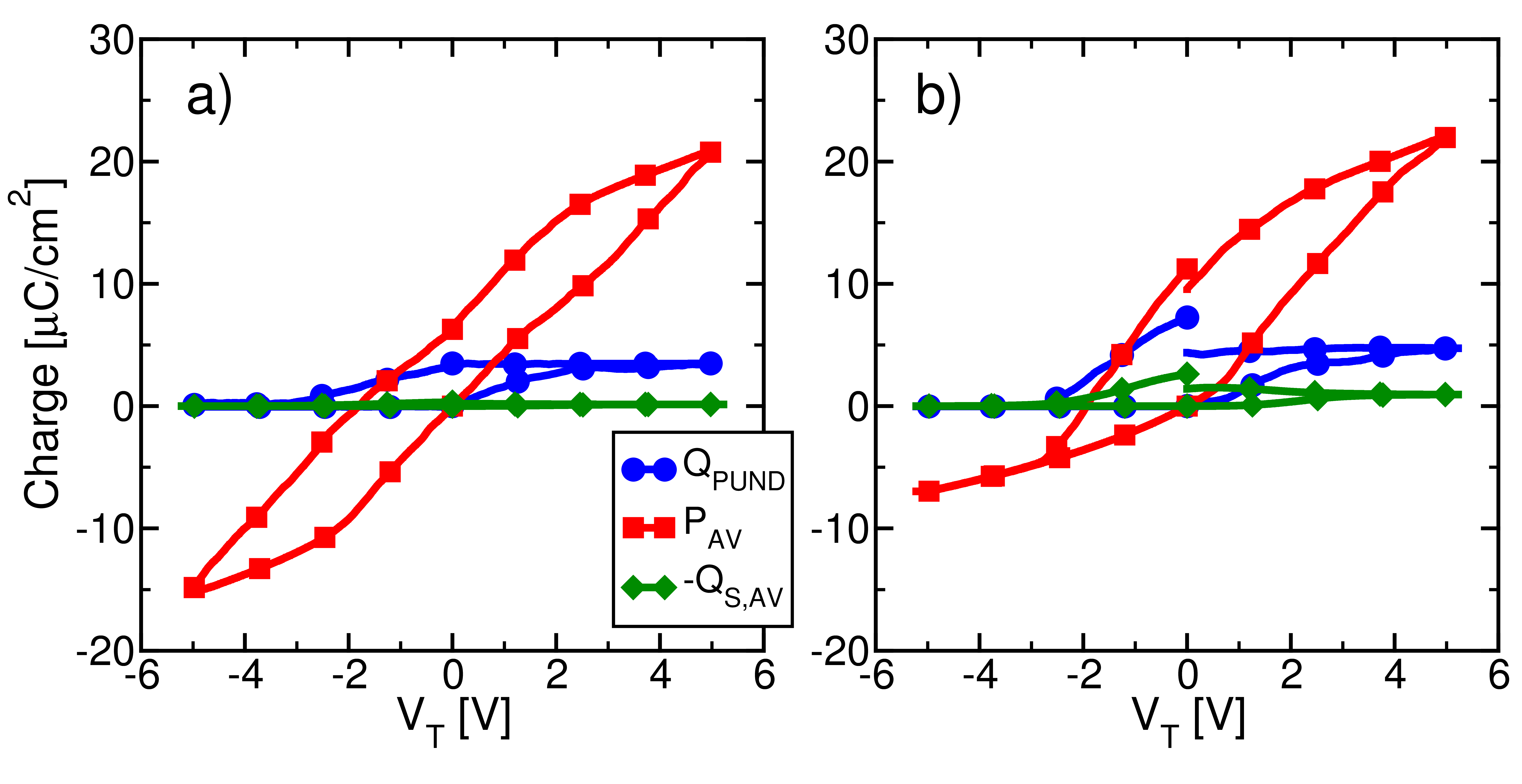}
	\caption{\footnotesize Same simulations for a PUND waveform in an MFDM structure as in Figure \ref{fig:pund_nacc}, but for an oxide thickness $t_D$ of 2.5 nm. a) Results for acceptor and donor type trap densities $N_{acc}$$=$$N_{don}$$=$$0.5\times 10^{13}$ [cm$^{-2}$eV$^{-1}$]; b) Results for $N_{acc}$$=$$N_{don}$$=$$4\times 10^{13}$ [cm$^{-2}$eV$^{-1}$].
	}
	\label{fig:high_tD}
	\vspace{-0.2truecm}
\end{figure}

In Figure \ref{fig:pund_nacc} we report simulation results for $t_D = $ 1.5 nm and for different trap densities. The $Q_{PUND}$ is here defined as either $Q_{PU}$$=$$(Q_P-Q_U)$ or $Q_{ND}$$=$$(Q_N-Q_D)$, respectively for the positive and negative $V_T$ values.
The sign of the interface charge $Q_{S,AV}$ is typically opposite to the sign of the polarization, and Figure \ref{fig:pund_nacc} reports $-Q_{S,AV}$, which together with $P_{AV}$ determines $E_{DP,AV}$ according to Equation \ref{Eq:Edep_Pav_QSav}.
All charges in Figure \ref{fig:pund_nacc} are referred to the corresponding value at the beginning of the P pulse, namely at $t$$=$0 s and $V_T$$=$0 V in Figure \ref{Fig:PUND_sketch} (see also the discussion about Equations \ref{Eq:QP} and \ref{Eq:QP_QU}).
The $Q_{PUND}$ in Figure \ref{fig:pund_nacc}(a) shows a hysteresis loop that is much more tilted and stretched than in the corresponding MFM curves (filled triangles). The features for an MFDM are similar to those experimentally observed in the $P$-$V$ curves for an HZO capacitor serially connected to a discrete ceramic capacitor ensuring a negligible charge injection,$^{\text{\cite{Park_Nanoscale2021}}}$ or to measurements in MFDM structures with thicker Al$_2$O$_3$ layers.$^{\text{\cite{Hoffmann_AdvFunctionalMaterials_2021}}}$ In fact, the relatively low density of traps in Figure \ref{fig:pund_nacc}(a) results in an interface charge $Q_S$ (green diamonds) that is practically negligible compared to the ferroelectric polarization (red squares). The lack of any compensation of the polarization results in a large depolarization field $E_{DP,AV}$ (see Equation \ref{Eq:Edep_Pav_QSav}), which in turn leads to a vast discrepancy between $Q_{PUND}$ and $P_{AV}$. In fact, Figure \ref{fig:pund_nacc}(a) shows that for $|V_T|$ above about 4 V a complete polarization switching occurs. Nevertheless the corresponding $Q_{PUND}$ is much smaller than $P_{AV}$, mainly because the $E_{DP,AV}$ term in Equation \ref{Eq:QP_QU} subtracts from the $P_{AV}$ term due to the opposite sign.
The results for $Q_{PUND}$ are quite different in Figure \ref{fig:pund_nacc}(b), because the $Q_{S,AV}$ can now compensate $P_{AV}$ to a large extent, thus drastically reducing the depolarization field. The hysteresis loop of the $Q_{PUND}$ curve in Figure \ref{fig:pund_nacc}(b) is qualitatively similar to the experimental behaviour observed in FTJ structures with a thin tunnel oxide,$^{\text{\cite{Fontanini_ESSDERC2021,Li_VLSI_2020}}}$ and the discrepancy between $Q_{PUND}$ and $P_{AV}$ is much smaller than in Figure \ref{fig:pund_nacc}(a).
Figure \ref{fig:high_tD} illustrates the same analysis as in Figure \ref{fig:pund_nacc}, but for a larger dielectric thickness $t_D$ = $2.5$ nm. In this case the results of both small and large interface traps densities have a qualitative behaviour similar to Figure \ref{fig:pund_nacc}(a), namely the $Q_S$ is very small (green diamonds) and the compensation of the ferroelectric polarization is minimal. For both traps densities, the depolarization field results in a $Q_{PUND}$ much smaller than $P_{AV}$.
The different behaviour in Figure \ref{fig:high_tD}(b) compared to Figure \ref{fig:pund_nacc}(b) is due to the fact that, according to the tunneling effective masses and traps cross-sections reported in Table \ref{Tab:Sims_Param}, the trapping and de-trapping dynamics cannot  follow the 1 kHz $V_T$ waveform for $t_D$$=$2.5 nm or larger. In fact, because the HZO layer is 10 nm thick, in the simulations of this work the trapping dynamics is essentially set by the tunnelling through the much thinner dielectric layer. The lack of $Q_S$ modulation in Figure \ref{fig:high_tD}(b) is thus a dynamic effect. This emphasizes that the trapping induced compensation of the ferroelectric polarization requires both a large enough trap density at the FE-DE interface, and a trapping dynamics fast enough to respond to the $V_T$ waveform. This latter observation has been crucial in transient negative capacitance experiments, where thick dielectrics and fast bias waveforms were used to avoid the undesired compensation of the ferroelectric polarization and to achieve a hysteresis free behaviour.$^{\text{\cite{Rollo_Nanoscale2020,Hoffmann_Nature2019,Esseni_Nanoscale2021}}}$

While the $t_D$ values at which traps can no longer respond to a given $V_T$ waveform depend on the tunnelling model and the corresponding parameters in Table \ref{Tab:Sims_Param}, the qualitative trend is expected to be independent of the modelling details.

\begin{figure}[!h]
	\centering	\includegraphics[width=0.69\linewidth]{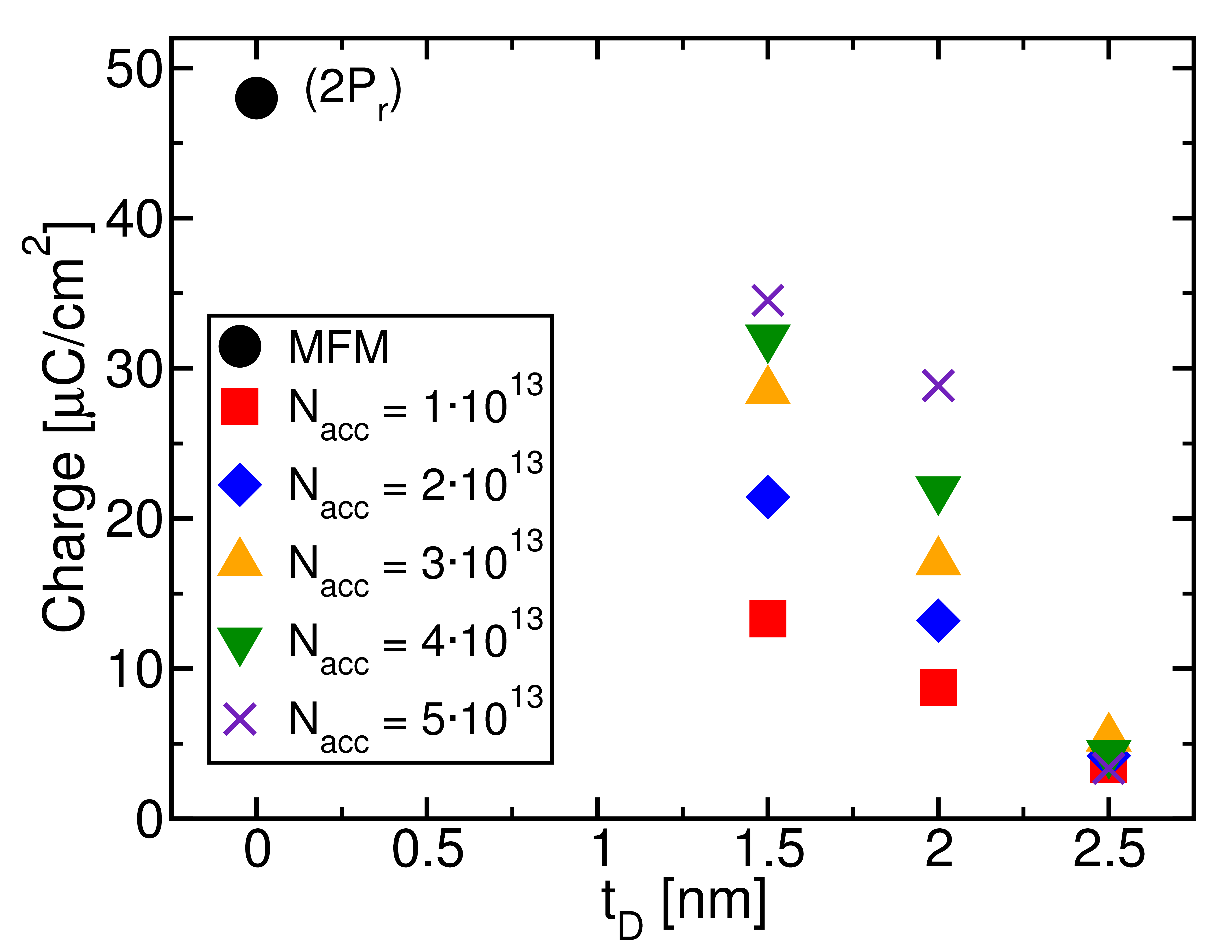}
	\caption{$Q_{PU}$ extracted from PUND simulations in MFDM structures for $V_T $ = 0 V, for different dielectric thickness $t_D$ and different traps density $N_{acc}$$=$$N_{don}$ (in units of cm$^{-2}$eV$^{-1}$). The remnant polarization $2P_r$ extracted for an MFM structure is also reported for comparison.}
	\label{Fig:2pr}
	\vspace{-0.4truecm}
\end{figure}

\subsection{Discrepancies between $Q_{PUND}$ and $P_{AV}$}
\label{Sec:Errors}
The discrepancies between $Q_{PUND}$ and $P_{AV}$ shown in Figures \ref{fig:pund_nacc} and \ref{fig:high_tD} correspond to errors in the outcome of the PUND method. Hence in this section we evaluate the relative error $E_{PU}$$=$$|Q_{PU}-P_{AV}^{(P)}|$$/$$P_{AV}^{(P)}$,
where from hereon all the quantities are evaluated at the end of the P or the U pulse, namely when $V_T$ is zero.
Similar definitions apply to the N and D pulses, and the simulation results are also completely similar (not shown).
\begin{figure}[htp]
	\centering
	\includegraphics[width=0.8\linewidth]{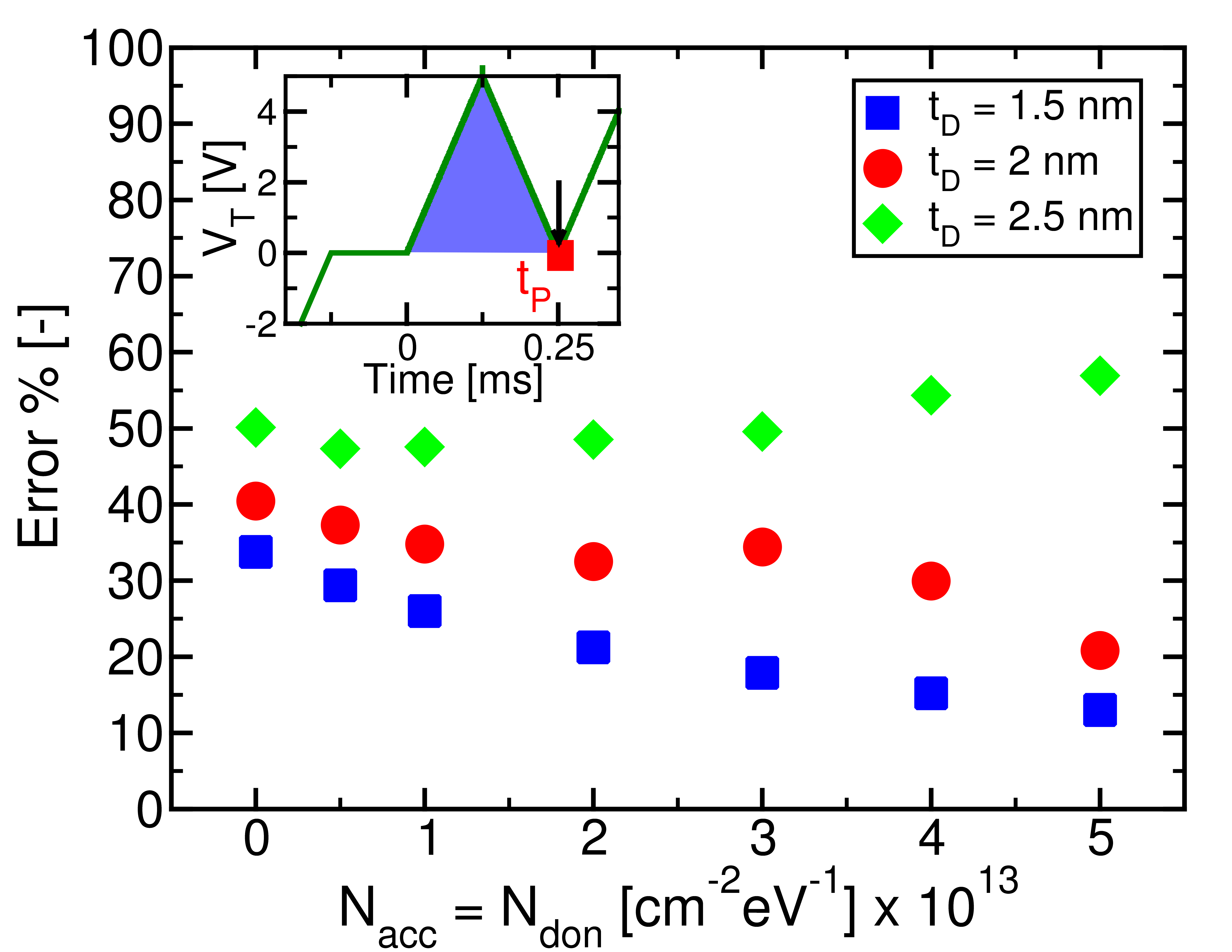}
	\caption{\footnotesize Error $|Q_{PU}-P_{AV}^{(P)}|$$/$$P_{AV}^{(P)}$ of PUND measurements in an MFDM structure
	for different thicknesses $t_D$ and different trap densities $N_{acc}$ = $N_{don}$. The error is calculated for the P and U pulses. a) Error evaluated at the end of the P pulse (see inset);  b) Error evaluated at the peak of the P pulse pulse.}
	\label{fig:errors}
	\vspace{-0.1truecm}
\end{figure}

\begin{figure}[htb]
	\vspace{-0.4truecm}
\centering
\includegraphics[width=1\linewidth]{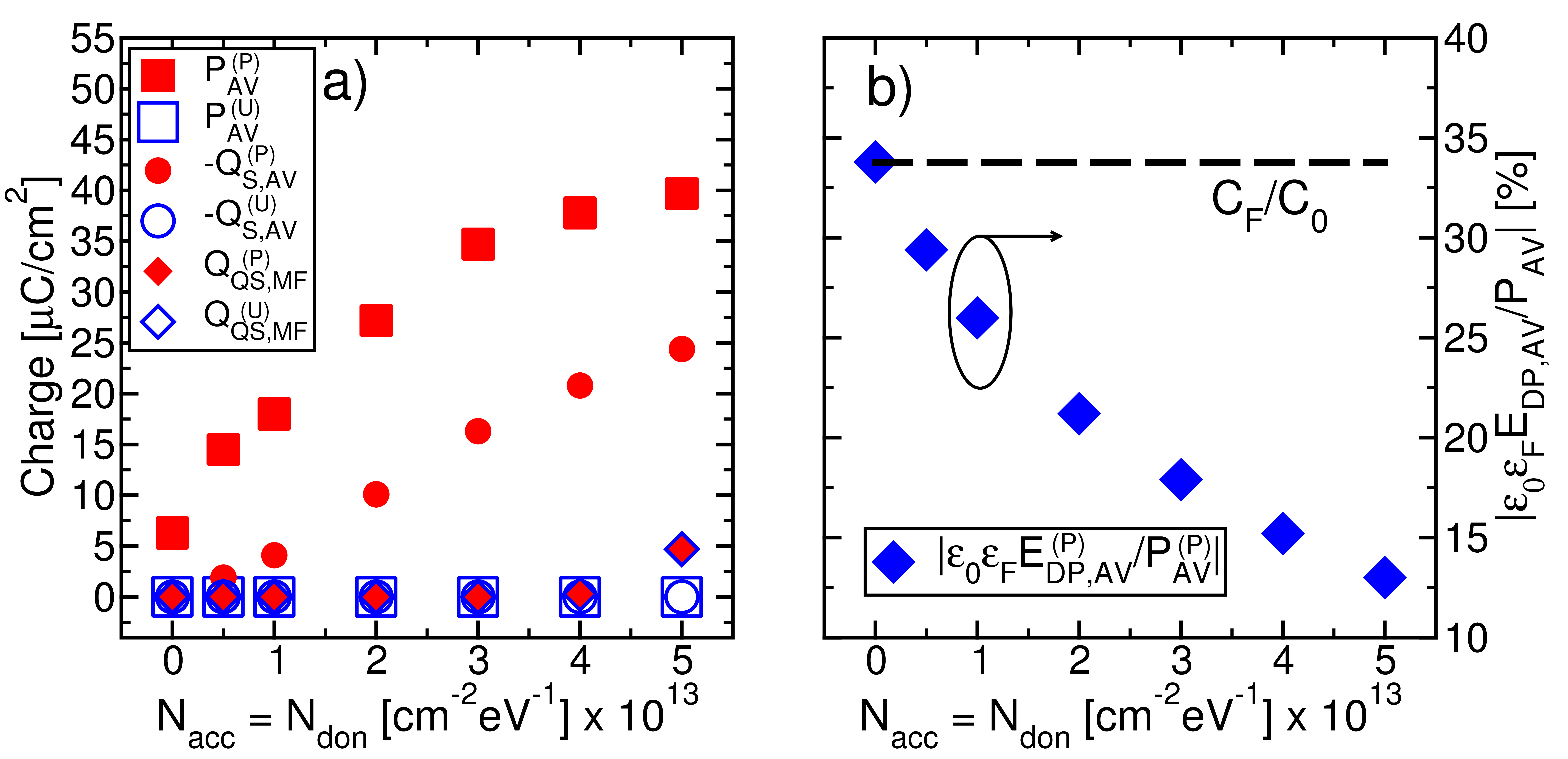}
\caption{\footnotesize Charge components contributing to $Q_{PU}$ according to Equation \ref{Eq:QP_QU_2} and Equation \ref{Eq:QP_QU} evaluated for $t_D$$=$1.5 nm and in the same conditions as in Figure \ref{fig:errors} (i.e. at the end of the P pulse with $V_T$ = 0 V). The sign of $Q_{S,AV}$ and $\varepsilon_0 \varepsilon_FE_{DP,AV}^{(P)}$ is opposite to the sign of $P_{AV}$ and the figure displays $-$$Q_S$ and  $-\varepsilon_0 \varepsilon_FE_{DP,AV}^{(P)}$. All the quantities shown in the figure are difference between the values at the end and at the start of the P pulse.}
\label{fig:components}
\end{figure}

Figure \ref{Fig:2pr} shows the $Q_{PU}$ of the MFDM structures. It can be seen that a combination of a large $t_D$ and low concentrations of traps lead to low simulated $Q_{PU}$ values, because the corresponding  $\varepsilon_0\varepsilon_{F} E_{DP,AV}^{(P)}$ term is comparable to $P_{AV}^{(P)}$ (as later shown in Figure \ref{fig:components}).
Figure \ref{fig:errors} reports the evaluation of the error $E_{PU}$ for different dielectric thicknesses and trap densities. As it can be seen, the error tends to decrease for increasing trap densities, due to the corresponding reduction of the depolarization field $E_{DP}$. For the same reason the error increases for thicker dielectrics. This latter behavior results in a $t_D$ dependence of the $P_{AV}$ estimated by the PUND method, which is an artifact of the method when it is applied to an MFDM structure.\\
For t$_D$ = 2.5 nm the error is fairly insensitive to the trap density, because the $Q_S$ in the traps cannot respond to the $V_T$ waveform according to our tunnelling model.
To gain an insight about the main causes of the errors shown in Figure \ref{fig:errors}, we first rewrite Equation \ref{Eq:QP_QU} as
\begin{equation}
	\begin{split}
		Q_{PU} \approx &\dfrac{C_D}{C_0} \left ( P_{AV}^{(P)} - P_{AV}^{(U)} \right ) - \dfrac{C_F}{C_0} \left ( Q_{S,AV}^{(P)} - Q_{S,AV}^{(U)}\right ) + \\
		&+Q_{QS,MF}^{(P)}-Q_{QS,MF}^{(U)}
	\end{split}
\label{Eq:QP_QU_2}
\end{equation}
where we have used Equation \ref{Eq:Edep_Pav_QSav} to express the depolarization field $E_{DP,AV}^{(P)}$; here we have omitted the leakage part because the leakage current is not included in our simulations, and all the quantities in Equation \ref{Eq:QP_QU_2} are evaluated at the end of the P or the U pulse.
Then we report in Figure \ref{fig:components}(a) the quantities in the right hand side of Equations \ref{Eq:Edep_Pav_QSav}, \ref{Eq:QP_QU} and \ref{Eq:QP_QU_2}, for a dielectric thickness $t_D$ = 1.5 nm and evaluated in the same condition used to evaluate the PUND error in Figure \ref{fig:errors} (i.e. $V_T=0$ V).\\
Figure \ref{fig:components}(a) conveys several important messages. The terms $Q_{QS,MF}^{(P)}$, $Q_{QS,MF}^{(U)}$ (diamonds) related to the trapping and de-trapping current at the MF electrode are very small even for large trap densities, hence they do not appreciably influence $Q_{PU}$ in Equations \ref{Eq:QP_QU} and \ref{Eq:QP_QU_2}. This is not surprising because, in the MFDM structures at study, traps exchange electrons primarily with the MD electrode as the dielectric is much thinner than the ferroelectric layer.\\
Moreover, at large trap densities the $Q_{S,AV}^{(P)}$ (filled circles) in the P pulse is comparable to $P_{AV}^{(P)}$ (filled squares), whereas $Q_{S,AV}^{(P)}$ becomes negligible at low trap densities. The $Q_{S,AV}^{(U)}$ in the $U$ pulse, instead, is always negligible compared to $P_{AV}^{(P)}$.
This is because, for the case at study in Figure \ref{fig:components}, the band bending in the dielectric at the end of the P pulse is such that the energy levels of both acceptor and donor traps fall below the Fermi level of the MD contact (see Figure \ref{Fig:Charges_Currents}(b)). Hence, essentially all traps have been filled at the end of the P pulse, and their occupation is not appreciably changed during the following U pulse. Figure \ref{fig:components}(a) shows that also $P_{AV}^{(U)}$ in the $U$ pulse is much smaller than $P_{AV}^{(P)}$. This is because the $P_{AV}$ in the P and U pulse is a measure of the non reversible switching, whereas most of the switching in the $U$ pulse is reversible in nature because it is the switching of those domains that have back switched after the $P$ pulse.\\
As mentioned above, Figure \ref{fig:components}(a) shows that at low trap densities we have $|Q_{S,AV}^{(P)}|$$\ll$$P_{AV}^{(P)}$ and Equation \ref{Eq:Edep_Pav_QSav} suggests that this results in a $\varepsilon_0\varepsilon_{F} \, E_{DP,AV}^{(P)}$$\simeq$$-(C_F/C_0)P_{AV}^{(P)}$, as it is confirmed by Figure \ref{fig:components}(b). These are the conditions that in Figure \ref{fig:errors} correspond to the maximum discrepancy between $Q_{PU}$ and $P_{AV}^{(P)}$. Equation \ref{Eq:QP_QU_2}  shows that for $|Q_{S,AV}^{(P)}|$$\ll$$P_{AV}^{(P)}$ the extracted $Q_{PU}$ tends to $(C_D/C_0)P_{AV}^{(P)}$, in fact resulting in a large underestimate of $P_{AV}^{(P)}$. At large trap densities, instead, $|Q_{S,AV}^{(P)}|$ becomes comparable to $P_{AV}^{(P)}$ and Equation \ref{Eq:Edep_Pav_QSav} predicts a drastic reduction of the  $|\varepsilon_0\varepsilon_{F} \, E_{DP,AV}^{(P)}/P_{AV}^{(P)}|$ term, which can be observed in Figure \ref{fig:components}(b). The error in Figure \ref{fig:errors} is correspondingly reduced at large trap densities; in fact Equation \ref{Eq:QP_QU} suggests that $Q_{PU}$ tends to $P_{AV}^{(P)}$.

\vspace{-0.6truecm}

\section{Conclusions}
\label{Sec:Conclusions}

We have revisited the theory and application of the PUND technique in MFDM structures by using analytical derivations and numerical simulations. The interplay between the depolarization field and charge trapping in an MFDM stack makes it difficult to obtain from the terminal currents alone an accurate estimate of the spontaneous polarization switched in the $P$ or in the $N$ pulse.

The discrepancies between $Q_{PU}$ and $P_{AV}^{(P)}$, for example, were analyzed for different thicknesses $t_D$ of the dielectric layer and different traps densities at the FE-DE interface, that in turn result in different trapping induced compensations of the ferroelectric polarization. Because in simulations one can inspect all the physical quantities at play, even those that are not usually accessible in experiments, our analysis allowed us to gain an insight about the main sources of error for the PUND technique in MFDM structures.
Besides the discrepancies between $Q_{PU}$ and $P_{AV}^{(P)}$ that can be identified as an error of the PUND technique, it should be understood that neither the $Q_{PU}$ nor the $P_{AV}^{(P)}$ of an MFDM structure are a good estimate of the $2P_r$ of the underlying ferroelectric. This is because the depolarization field can be large at zero external bias, so that the MFDM structure at $V_T$$=$0 is not at all representative of the ferroelectric material at zero ferroelectric field. More precisely the $Q_{PU}$ of the PUND technique tends to underestimate the non-reversible switched polarization $P_{AV}^{(P)}$, which in turn is an underestimate of the $2P_r$ of the ferroelectric. The differences between these quantities depend on $t_D$ and on the density of traps, which may lead to artifacts in the characterization of a possible $t_D$ dependence of the properties of the underlying ferroelectric layer. Of course, we acknowledge that it would be very useful to suggest corrections to the PUND technique or to propose a novel technique for MFDM structures in order to duly account for the depolarizing field, and maybe even separate the switched polarization from the trapped charge. At the time of writing, however, we are not able to suggest a clear way of achieving such targets in an MFDM structure, and by relying exclusively on quantities accessible in experiments. In this respect, we cannot but conclude that more work is needed to improve the electrical probing of spontaneous polarization in
ferroelectric-dielectric heterostructures.

\section*{Acknowledgements}
This work was supported by European Union through the BeFerroSynaptic project (GA:871737).

{\appendices
\section{Green's function of a point charge in the MFDM stack}
\label{Sec:SUPPL_Eq_Pav}
\begin{figure}[bp]
	\centering
	\includegraphics[width=0.9\linewidth]{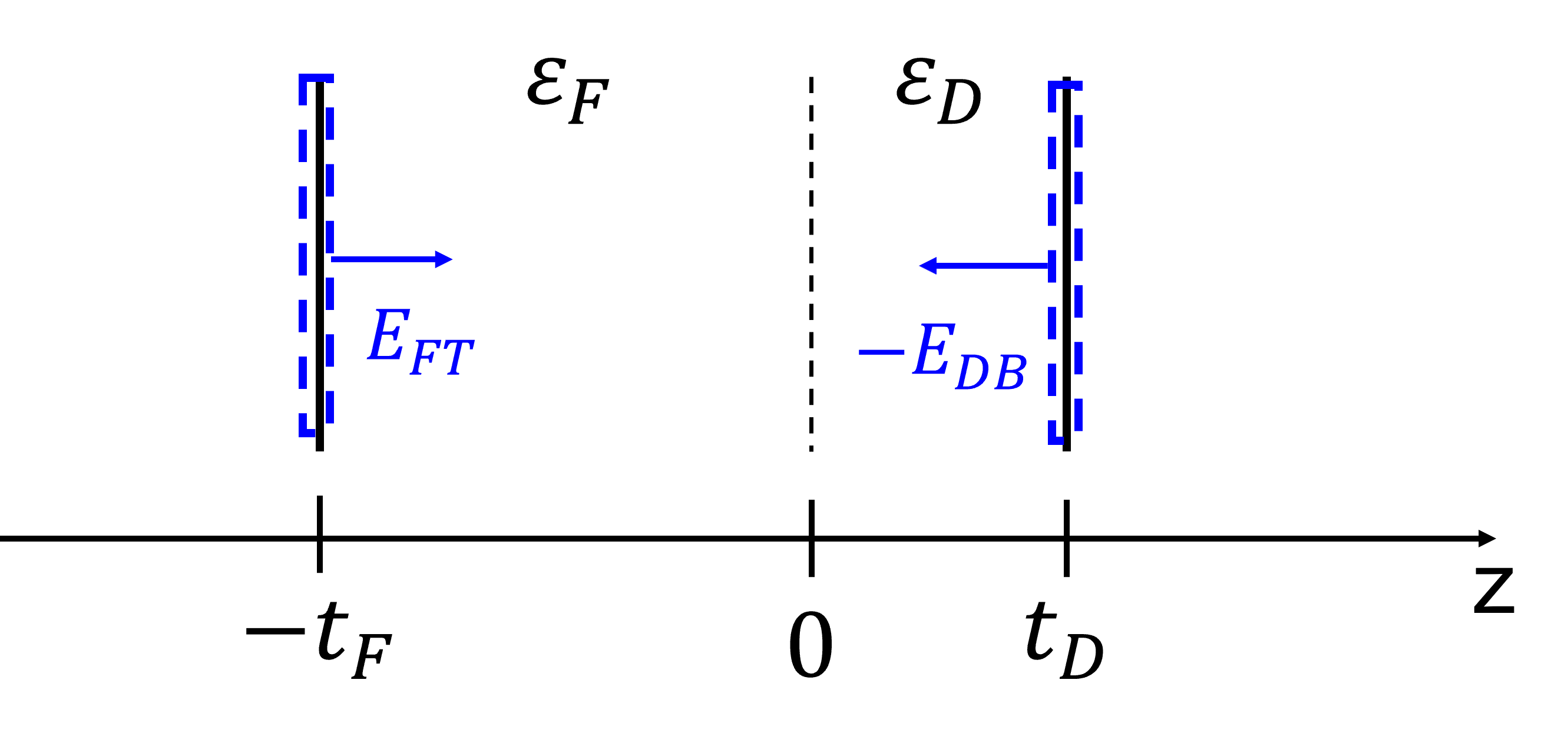}
	\caption{\footnotesize Sketch of the MFDM stack, where $E_{FT}$ and $E_{DB}$ denote the $z$ component of the electric field respectively at MF-FE interface and at MD-DE interface.}
	\label{Fig:sketchsuppinfo}
\end{figure}

In this section we discuss the analytical expression for the Green's function of the point charge defined in Equation \ref{Eq:GMF_DEF}.
The potential $\psi(\textbf{r},z)$ produced by a point charge located in $(\textbf{r}_0,z_0)$ in a dielectric material having a relative dielectric constant $\varepsilon_r$ can be obtained by solving the Poisson equation$^{\text{\cite{NanoscaleMOS_Cambridge2011}}}$
\begin{equation}
	\nabla^2\psi(\textbf{r},z) = -\dfrac{e}{\varepsilon_0 \varepsilon_r}\delta (\textbf{r}-\textbf{r}_0)\delta (z-z_0).
	\label{Eq:S_poisson}
\end{equation}
where $e$ is the elementary charge. We now introduce the 2D Fourier transform of $\psi(\textbf{r},z)$ with respect to the coordinates \rv$=$($x$,$y$), and define the Fourier pair
\begin{equation}
	\begin{split}
	\psi(\textbf{r},z) &\approx \int_{\qv} \psi(\qv,z) \, \expr{- i \, \qv \cdot {\bf r}} \, d\qv \\
	\psi(\qv,z) &\approx \frac{1}{(2 \pi)^2}
	\int_{A} \psi(\textbf{r},z) \, \expr{i \, \qv \cdot {\bf r}} \, d\rv  \,  \,  .
	\end{split}
	\label{Eq:FTransf_2D}
\end{equation}
Equation \ref{Eq:FTransf_2D} assumes that the device area $A$ is large enough that the integral over $A$ is a good approximation of the indefinite integral over the entire ($x$,$y$) plane. \footnote{The formalism may be rephrased in terms of a Fourier series by assuming periodic boundary conditions for $\psi(\textbf{r},z)$ at the edges of the area $A$, that would however lead to identical results.$^{\text{\cite{NanoscaleMOS_Cambridge2011}}}$}

\noindent
If we now recall the identity
\begin{equation}
	\delta (\textbf{r}-\textbf{r}_0)  \approx \dfrac{1}{(2\pi)^2}\int_{\qv}\exps{-i\,\qv\cdot(\rv-\rvz)} \, d\qv
	\label{Eq:delta}
\end{equation}
and substitute Equations \ref{Eq:FTransf_2D}, \ref{Eq:delta} into Equation \ref{Eq:S_poisson}, we can readily infer that the unknown potential $\Psi(\qv,z)$
takes the form$^{\text{\cite{NanoscaleMOS_Cambridge2011}}}$
\begin{equation}
	\psi(\qv,z) =
	\dfrac{\expr{i\,\qv\cdot\rvz}}{(2\pi)^2} \, \phi(q,z)
	\label{Eq:Psi_r0_2pi}
\end{equation}
where $\phi(q,z)$ must satisfy the differential equation
\begin{equation}
	\label{identically} \bigg[\frac{\partial^2}{\partial
		z^2}-q^2\bigg] \, \phi = -\frac{e}{\varepsilon_0 \varepsilon_r} \, \, \delta(z-z_0)
\end{equation}

\noindent
Let us now assume that the point charge is located at $z_0$$=$0, namely at  the FE-DE interface (see Figure \ref{Fig:sketchsuppinfo}). In this case the potential $\phi_F(q,z)$ in the ferroelectric and $\phi_D(q,z)$ in the dielectric region can be written as
\begin{subequations}
	\begin{align}
		\phi_F (z) &= C_1\,\expr{q\, z} + C_2\,\expr{-q\, z} \quad\quad z<0 \label{Eq:S_phiF_general}\\
		\phi_D (z) &= C_3\,\expr{q\, z} + C_4\,\expr{-q\, z} \quad \quad z>0 \label{Eq:S_phiD_general}
	\end{align}
\end{subequations}
where the four $q$ dependent constants $C_1$, $C_2$, $C_3$ and $C_4$ can be determined by using appropriate boundary conditions. At the interface with metal electrodes we used $\phi_F(-t_F)$$=$0, $\phi_D(t_D)$$=$0 , whereas at the FE-DE interface we employed the conditions $\phi_F(0)$$=$$\phi_D(0)$ and $[\varepsilon_0 \varepsilon_F (\partial \phi_F(0)/\partial z)-\varepsilon_0 \varepsilon_D (\partial \phi_D(0)/\partial z)]$$=$$e$.
By doing so we obtain
\begin{subequations}
	\begin{align}
		\phi_F(z) &= C_{Fe}\,\left\{ \expr{q\, z} - \exps{-q\left( z+2\, t_F \right)} \right\} \quad z<0 \label{Eq:S_phiF_final}\\
		\phi_D(z) &= C_{De} \, \left\{ \expr{q\,z} - \exps{-q\left(z-2\,t_D\right)} \right\} \quad z>0 \label{Eq:S_phiD_final}
	\end{align}
\end{subequations}
with
\begin{subequations}
	\begin{align}
		C_{Fe} &= \dfrac{1-\text{e}_D}{\varepsilon_F\left( 1 + \text{e}_F \right)\left( 1-\text{e}_D \right)-\varepsilon_{D}\left( 1 - \text{e}_F \right)\left( 1+\text{e}_D \right)}\cdot\dfrac{e}{q} \label{Eq:C1_express} \\
		C_{De} &= \dfrac{1-\text{e}_F}{ 1-\text{e}_D )} \cdot \, C_{Fe} \label{Eq:C3_express}
	\end{align}
\end{subequations}
where we have the notation more compact by introducing $\text{e}_F$$=$$\expr{-2\, q\, t_F}$ and $\text{e}_D$$=$$\expr{2\, q\, t_D}$.
The Green's function $G_{MF}(\textbf{r}_0,z_0)$ that we wish to determine is defined as
\begin{equation}
	G_{MF}(\textbf{r}_0,z_0) = \dfrac{\varepsilon_0 \varepsilon_F}{e} \int\displaylimits_A E_{FT}(\textbf{r})\,d \textbf{r}  \, \,
	\label{Eq:GMF_APP}
\end{equation}
where
$E_{FT}(\textbf{r})$ denotes the $z$ component of the electric field at the MF-FE interface (i.e. at $z$$=$$-t_F$) produced by a point charge $e$ located at $(\textbf{r}_0,z_0)$.
By recalling the definition of the Fourier transform pairs in Equation \ref{Eq:FTransf_2D} and then using Equation \ref{Eq:Psi_r0_2pi}, we have
\begin{equation}
	\begin{split}
		\int\displaylimits_A E_{FT}(\textbf{r})\,d \textbf{r}  &=   (2 \pi)^2 \, \lim_{q\to 0} E_{FT}(\qv) = \\
		&= - \, (2 \pi)^2 \, \lim_{q\to 0} \dfrac{\partial \psi(\qv,z_0)}{\partial z} = \\
		&= - \, \lim_{q\to 0}  \dfrac{\partial \phi_F(\qv,z_0)}{\partial z}
	\end{split}
	\label{Eq:IntR_LimitK}
\end{equation}
Equations\ref{Eq:GMF_APP}, \ref{Eq:IntR_LimitK} finally provide
\begin{equation}
	G_{MF}(\textbf{r}_0,z_0) = - \dfrac{\varepsilon_0 \varepsilon_F}{e} \, \lim_{q\to 0}  \dfrac{\partial \phi_F(\qv,z_0)}{\partial z}   \, \, .
	\label{Eq:GMF_calcul}
\end{equation}

\noindent
For $z_0$$=$$-t_F$ the limit in Equation\ref{Eq:GMF_calcul} can be readily calculated by using Equations \ref{Eq:S_phiF_final} and \ref{Eq:C1_express}, so as to obtain
\begin{equation}
	G_{MF}(\textbf{r}_0,0) = -\dfrac{C_F}{C_F+C_D} \label{Eq:S_GMF}
\end{equation}

\noindent
The corresponding Green's function $G_{MD}$ at the MD electrode, defined in Equation \ref{Eq:GMD_DEF}, can be derived with an entirely similar procedure. The result is
\begin{equation}
	G_{MD}(\textbf{r}_0,0) = - \dfrac{\varepsilon_0 \varepsilon_D}{e} \, \lim_{q\to 0}  \dfrac{\partial \phi_D(\qv,t_D)}{\partial z} =  -\dfrac{C_D}{C_F+C_D}
	\label{Eq:S_GMD}
\end{equation}
so that $(G_{MF}(\textbf{r}_0,0)+G_{MD}(\textbf{r}_0,0))$$=$$-$1.

Similar derivations apply to the case of a point charge located in the ferroelectric (i.e. for $-t_F<z_0<0$) or in the dielectric (i.e. for $0<z_0<t_d$). In the former case we obtain
\begin{align}
	G_{MF}(\textbf{r}_0,z_0) &= -\dfrac{C_F+C_D\dfrac{\lvert z_0\rvert}{t_F}}{C_F+C_D} \label{Eq:S_GMF_fe}\\
	\nonumber \\
	G_{MD}(\textbf{r}_0,z_0) &= -\dfrac{C_D\, \left( 1-\dfrac{\lvert z_0 \rvert}{t_F} \right)}{C_F+C_D} \label{Eq:S_GMD_fe}
\end{align}
whereas in the latter case we have
\begin{align}
	G_{MF}(\textbf{r}_0,z_0) & = -\dfrac{C_F\, \left( 1-\dfrac{z_0}{t_D} \right)}{C_F + C_D} \label{Eq:S_GMF_de}\\
	\nonumber \\
	G_{MD}(\textbf{r}_0,z_0) &= -\dfrac{C_D + C_F \dfrac{z_0}{t_D}}{C_F+C_D}. \label{Eq:S_GMD_de}
\end{align}

\noindent
As it can be seen, even for an $z_0$$\ne$0 we have $[G_{MF}(\textbf{r}_0,z_0) + G_{MD}(\textbf{r}_0,z_0)] = -1$.
\section{Charge and current at the MD electrode}
\label{Sec:Qmd_suppl}

The analysis of PUND measurements presented in the main paper is based on the current $I_{MF}$ at the MF electrode. According to the $I_{MF}$, $I_{MD}$ definitions sketched in Figure \ref{Fig:MFIM_Sketches}, we see that $I_{MD}$ must be equal to $I_{MF}$.
For the completeness of definitions and derivations, we here report a concise analysis about the charge $Q_{MD}$ and current $I_{MD}$ at the MD electrode.
We start with Q$_\text{MD}$ written as (see Figure \ref{Fig:sketchsuppinfo})
\begin{equation}
	Q_{MD}\, (t) = \dfrac{1}{A}\int\displaylimits_A - \varepsilon_0\varepsilon_{D}E_{DB}(\textbf{r},t)\, d\textbf{r} \, = - \varepsilon_0\varepsilon_{D}E_{DB,AV}(t)
	\label{Eq:Suppl_QMD}
\end{equation}
where $E_{DB}(\textbf{r},t)$ is the $z$ component of the field at the DE-MD interface at $z$$=$$t_D$. The term  $\varepsilon_0\varepsilon_{D}E_{DB}(\textbf{r})$ can be expressed as
\begin{equation}
	\varepsilon_0\varepsilon_{D}E_{DB}(\textbf{r}) = C_S\, V_T + \varepsilon_0\varepsilon_D E_{DI}(\textbf{r})
\end{equation}
where $E_{DI}(\textbf{r})$ is the contribution to the field due to the total charge $\left[ P(\textbf{r}_0)+Q_S(\textbf{r}_0) \right]$ at the FE-DE interface.
As already discussed in Section \ref{Sec:General_Model} of the main paper, we are here assuming that trapping is dominated by interface traps at the FE-DE interface.\\
We can now define the Green's function $G_{MD}(\textbf{r}_0,z_0)$ at the MD electrode
\begin{equation}
	G_{MD}(\textbf{r}_0,z_0) = \dfrac{\varepsilon_0\varepsilon_{D}}{e}\int\displaylimits_A -E_{DB}(\textbf{r},\textbf{r}_0z_0)\, d\textbf{r}
	\label{Eq:GMD_DEF}
\end{equation}
that allows us to write the average $E_{DI,AV}$ as
\begin{equation}
	\varepsilon_0\varepsilon_{D}E_{DI,AV} = -\dfrac{1}{A}\int\displaylimits_A\left[ P(\textbf{r}_0) + Q_S(\textbf{r}_0) \right] G_{MD}(\textbf{r}_0,z_0)\, d\textbf{r}_0
	\label{Eq:Suppl_EDIav}
\end{equation}
\noindent
For a charge at the FE-DE interface we have $G_{MD}(\textbf{r}_0,0) \simeq -C_D/C_0$ and thus
\begin{equation}
	\varepsilon_0\varepsilon_DE_{DI,AV} \simeq \dfrac{C_D}{C_0}\left(P_{AV} + Q_{S,AV}\right)
	\label{Eq:S_EDIav}
\end{equation}
so that $Q_{MD}(t)$ in Equation\ref{Eq:Suppl_QMD} becomes
\begin{equation}
	Q_{MD}(t) = -C_SV_T(t) -\varepsilon_0\varepsilon_DE_{DI,AV}(t).
	\label{Eq:S_QMD}
\end{equation}

\noindent
By using similar assumptions as those embraced in Section \ref{Sec:General_Model}, we can write $I_{MD}$ as
\begin{equation}
	\begin{split}
		I_{MD} &= -\dfrac{\partial Q_{MD}}{\partial t} + I_{QS,MD} + I_{lkg} \\
		& = C_S\dfrac{\partial V_T}{\partial t} + \varepsilon_0\varepsilon_D\dfrac{\partial E_{DI,AV}}{\partial t} +\\
		&+ I_{QS,MD} + I_{lkg}\\
		&=  C_S\dfrac{\partial V_T}{\partial t} + \dfrac{C_D}{C_0} \dfrac{\partial P_{AV}}{\partial t} + \dfrac{C_D}{C_0} \dfrac{\partial Q_{S,AV}}{\partial t} +\\
		&+ I_{QS,MD} + I_{lkg} \\
	\end{split}
\label{Eq:S_IMDgeneral}
\end{equation}
where in the last equality we have used Equation \ref{Eq:S_EDIav}.

\noindent
By recalling the $I_{MF}$ expression in Equation \ref{Eq:IMF_DEF} and the relation $[I_{QS,MF}-I_{QS,MD}]$$=$$\partial Q_{S,AV}/\partial t$,
we readily obtain $(I_{MF}$$-$$I_{MD})$$=$0, thus confirming that $I_{MD}$ is equal to $I_{MF}$.}

\vspace{-0.5truecm}

\bibliographystyle{./bibtex/MyIEEEtran.bst}
\bibliography{./bibtex/biblio}

%
%
%
%
%

\vfill

\end{document}